\documentclass[12pt,a4paper]{article}
\usepackage{amsmath}
\usepackage{amssymb}

\allowdisplaybreaks
\newcommand{\bs}{\boldsymbol}
\newcommand{\C}{\mathbb{C}}
\newcommand{\R}{\mathbb{R}}
\newcommand{\FS}{\mathbb{FS}}
\newcommand{\E}{\mathbf{E}}
\newcommand{\SL}{\mathop{\text{SL}}\nolimits}
\newcommand{\Herm}{\mathop{\text{Herm}}\nolimits}
\newcommand{\contr}{\mathop{\text{contraction}}\nolimits}
\newcommand{\tr}{\mathop{\text{trace}}\nolimits}
\newcommand{\valency}[4]{\genfrac{[}{.}{0pt}{1}{#1}{#3}%
\genfrac{.}{]}{0pt}{1}{#2}{#4}}

\begin{document}
\title{\textsf{Finslerian $N$-spinors: Algebra}}
\author{\textbf{A. V. Solov'yov}\thanks{Division of Theoretical Physics,
Faculty of Physics, Moscow State University, Moscow, Russia. E-mail:
anton@spin.phys.msu.su} \textbf{ and Yu. S. Vladimirov}\thanks{Division of
Theoretical Physics, Faculty of Physics, Moscow State University, Moscow,
Russia. E-mail: vlad@fund.phys.msu.su}}
\date{}
\maketitle

\begin{abstract}
\noindent
New mathematical objects called Finslerian $N$-spinors are discussed. The
Finslerian $N$-spinor algebra is developed. It is found that Finslerian
$N$-spinors are associated with an $N^2$-dimensional flat Finslerian space. A
generalization of the epimorphism $\SL(2,\C)\to\text{O}^\uparrow_+(1,3)$ to
a case of the group $\SL(N,\C)$ is constructed. Particular examples of
Finslerian $N$-spinors for $N=2,3$ are considered in detail.
\end{abstract}
\section*{1. Introduction}

Spinors as geometrical objects were discovered by \'E.~Cartan in 1913 (Cartan,
1913). One decade later, W.~Pauli (Pauli, 1927) and P.~A.~M.~Dirac (Dirac,
1928) rediscovered spinors in connection with the problem of describing the
spin of an electron. From that time, spinors are intensively used in
mathematics and physics.

In the classical works (Brauer and Weyl, 1935; Cartan, 1938), a concept of
the Cartan's 2-spinor was generalized and the theory of spinors in an arbitrary
$n$-dimensional pseudo-Euclidean space was constructed. In this article,
another generalization of 2-spinors is proposed which leads to the Finslerian
geometry. Originally, such a generalization appeared within the so-called {\it
relational theory of space-time\/} (Vladimirov, 1996; Solov'yov, 1996).
However, the corresponding mathematical scheme has also an independent meaning
and will be presented below.

In the next section, we shall develop a general algebraic formalism of
Finslerian $N$-spinors. The subsequent sections deal with the theory of the
simplest Finslerian 2- and 3-spinors. Conclusion contains some remarks
concerning the obtained results.

\section*{2. General formalism}

Let $\FS^N$ be a vector space of $N>1$ dimensions over $\C$ and
$$
[\cdot,\cdot,\dots,\cdot]\colon\underbrace{\FS^N\times\FS^N\times\cdots\times
\FS^N}_{\text{$N$ multiplicands}}\to\C\eqno(1)
$$
be a nonzero antisymmetric $N$-linear functional on $\FS^N$. The latter
means:\\
\indent
\phantom{ii}(i) there exist $\bs{\xi}_0$, $\bs{\eta}_0$, \dots, $\bs{\lambda}_0
\in\FS^N$ such that
$$
[\bs{\xi}_0,\bs{\eta}_0,\dots,\bs{\lambda}_0]=z_0\ne 0;\eqno(2)
$$
\indent
\phantom{i}(ii) for any $\bs{\xi}_1$, $\bs{\xi}_2$, \dots, $\bs{\xi}_N\in
\FS^N$,
$$
[\bs{\xi}_a,\bs{\xi}_b,\dots,\bs{\xi}_c]=\varepsilon_{ab\dots c}\,[\bs{\xi}_1,
\bs{\xi}_2,\dots,\bs{\xi}_N],
$$
where $a$, $b$, \dots, $c=1$, 2, \dots, $N$ and $\varepsilon_{ab\dots c}$ is
the $N$-dimensional Levi-Civita symbol with the ordinary normalization
$\varepsilon_{12\dots N}=1$;\\
\indent
(iii) for any $\bs{\xi}_1$, $\bs{\eta}_1$, $\bs{\xi}_2$, $\bs{\eta}_2$, \dots,
$\bs{\xi}_N$, $\bs{\eta}_N\in\FS^N$ and $z\in\C\,$,
\begin{align*}
[\bs{\xi}_1,\dots,\bs{\xi}_a+\bs{\eta}_a,\dots,\bs{\xi}_N]
&=
[\bs{\xi}_1,\dots,\bs{\xi}_a,\dots,\bs{\xi}_N]+[\bs{\xi}_1,\dots,\bs{\eta}_a,
\dots,\bs{\xi}_N],\\
[\bs{\xi}_1,\dots,z\bs{\xi}_a,\dots,\bs{\xi}_N]
&=
z[\bs{\xi}_1,\dots,\bs{\xi}_a,\dots,\bs{\xi}_N],
\end{align*}
where $a$ takes the values 1, 2, \dots, $N$.

We shall use the following terminology. The space $\FS^N$ equipped with
the functional (1) having the properties (i), (ii), and (iii) is called the
{\it space of Finslerian $N$-spinors}. The complex number $[\bs{\xi},
\bs{\eta},\dots,\bs{\lambda}]$ is respectively called the {\it scalar
$N$-product\/} of the Finslerian $N$-spinors $\bs{\xi}$, $\bs{\eta}$, \dots,
$\bs{\lambda}\in\FS^N$. 

It should be noted that $\bs{\xi}_0$, $\bs{\eta}_0$, \dots, $\bs{\lambda}_0$
are linearly independent. Indeed, if those were linearly dependent, one of the
Finslerian $N$-spinors $\bs{\xi}_0$, $\bs{\eta}_0$, \dots, $\bs{\lambda}_0$
would be a linear combination of the others and, in accordance with
(ii)--(iii), the scalar $N$-product $[\bs{\xi}_0,\bs{\eta}_0,\dots,
\bs{\lambda}_0]$ would be equal to zero. However, this is in contradiction with
(2). Thus, $\bs{\xi}_0$, $\bs{\eta}_0$, \dots, $\bs{\lambda}_0$ are linearly
independent, i.e., form a basis in $\FS^N$.

Let us introduce the notation $\bs{\epsilon}_1=\bs{\xi}_0$, $\bs{\epsilon}_2=
\bs{\eta}_0$, \dots, $\bs{\epsilon}_N=\bs{\lambda}_0/z_0$. It is evident that
the set $\{\bs{\epsilon}_1,\bs{\epsilon}_2,\dots,\bs{\epsilon}_N\}$ is a basis
in $\FS^N$. Due to (2) and (iii), its elements satisfy the condition
$$
[\bs{\epsilon}_1,\bs{\epsilon}_2,\dots,\bs{\epsilon}_N]=1.\eqno(3)
$$
We shall call such a basis {\it canonical}.

Let $\bs{\epsilon}_1^\prime$, $\bs{\epsilon}_2^\prime$, \dots,
$\bs{\epsilon}_N^\prime$ be arbitrary Finslerian $N$-spinors and
$$
\bs{\epsilon}_a^\prime=c_a^b\bs{\epsilon}_b\eqno(4)
$$
be their expansions into the canonical basis $\{\bs{\epsilon}_1,
\bs{\epsilon}_2,\dots,\bs{\epsilon}_N\}$; here $a$, $b=1$, 2, \dots, $N$,
$c_a^b\in\C\,$, and the summation is taken over the repeating index $b$.
With the help of (ii), (iii), (3), and (4), we find
$$
[\bs{\epsilon}_1^\prime,\bs{\epsilon}_2^\prime,\dots,\bs{\epsilon}_N^\prime]=
\det\text{C}_N,\eqno(5)
$$
where $\text{C}_N=\|c_a^b\|$. Since linear (in)dependence of
$\bs{\epsilon}_1^\prime$, $\bs{\epsilon}_2^\prime$, \dots,
$\bs{\epsilon}_N^\prime$ is equivalent to that of columns of the complex
$N{\times}N$ matrix $\text{C}_N$, the set $\{\bs{\epsilon}_1^\prime,
\bs{\epsilon}_2^\prime,\dots,\bs{\epsilon}_N^\prime\}$ is a basis in
$\FS^N$ if and only if $\det\text{C}_N\ne 0$. Moreover, it follows from
(5) that $\{\bs{\epsilon}_1^\prime,\bs{\epsilon}_2^\prime,\dots,
\bs{\epsilon}_N^\prime\}$ is a canonical one when $\det\text{C}_N=1$. Thus, if
$\text{C}_N$ runs the group $\SL(N,\C)$ of unimodular complex
$N{\times}N$ matrices, then $\{\bs{\epsilon}_1^\prime,\bs{\epsilon}_2^\prime,
\dots,\bs{\epsilon}_N^\prime\}$ runs the set $\E(\FS^N)$ of
canonical bases in $\FS^N$.

Let us express the scalar $N$-product of Finslerian $N$-spinors in terms of
their components with respect to {\it any\/} canonical basis $\{\bs{\epsilon}_1
,\dots,\bs{\epsilon}_N\}\in\E(\FS^N)$. By using (ii), (iii), (3), and the
expansions $\bs{\xi}=\xi^a\bs{\epsilon}_a$, $\bs{\eta}=\eta^b\bs{\epsilon}_b$,
\dots, $\bs{\lambda}=\lambda^c\bs{\epsilon}_c$, it is possible to show that
$$
[\bs{\xi},\bs{\eta},\dots,\bs{\lambda}]=\varepsilon_{ab\dots c}\,\xi^a\eta^b
\cdots\lambda^c,\eqno(6)
$$
where $\bs{\xi}$, $\bs{\eta}$, \dots, $\bs{\lambda}\in\FS^N$, $\xi^a$,
$\eta^b$, \dots, $\lambda^c\in\C\,$, $a$, $b$, \dots, $c=1$, 2, \dots,
$N$. In (6) as well as in the following formulas of this article, the summation
is taken over all the repeating indices. It is clear that the scalar
$N$-product (6) is zero if and only if $\bs{\xi}$, $\bs{\eta}$, \dots,
$\bs{\lambda}$ are linearly dependent Finslerian $N$-spinors.

Let us consider a mapping
\begin{align}
\bs{S}\colon\E(\FS^N)&\to\C\,^{N^{k+l+m+n}},\nonumber\\
\{\bs{\epsilon}_1,\dots,\bs{\epsilon}_N\}&\mapsto\bs{S}\{\bs{\epsilon}_1,\dots,
\bs{\epsilon}_N\}=\left(S_{a_1\!\dots a_m\dot{d}_1\!\dots\dot{d}_n}^{b_1\!\dots
b_k\dot{c}_1\!\dots\dot{c}_l}\!\{\bs{\epsilon}_1,\dots,\bs{\epsilon}_N\}\right)
\tag{7}
\end{align}
such that
\begin{align}
S_{a_1\!\dots a_m\dot{d}_1\!\dots\dot{d}_n}^{b_1\!\dots b_k\dot{c}_1\!\dots
\dot{c}_l}\!\{\bs{\epsilon}_1^\prime,\dots,\bs{\epsilon}_N^\prime\}
&=
c_{a_1}^{e_1}\!\cdots c_{a_m}^{e_m}
\overline{c_{\dot{d}_1}^{\dot{h}_1}}\!\cdots
\overline{c_{\dot{d}_n}^{\dot{h}_n}}
d^{b_1}_{f_1}\!\cdots d^{b_k}_{f_k}
\overline{d^{\dot{c}_1}_{\dot{g}_1}}\!\cdots
\overline{d^{\dot{c}_l}_{\dot{g}_l}}\nonumber\\
&\times
S_{e_1\!\dots e_m\dot{h}_1\!\dots\dot{h}_n}^{f_1\!\dots f_k\dot{g}_1\!\dots
\dot{g}_l}\!\{\bs{\epsilon}_1,\dots,\bs{\epsilon}_N\}\tag{8}
\end{align}
for any two canonical bases $\{\bs{\epsilon}_1,\dots,\bs{\epsilon}_N\}$,
$\{\bs{\epsilon}_1^\prime,\dots,\bs{\epsilon}_N^\prime\}\in\E(\FS^N)$ whose
elements are connected by the relations (4). Here all the indices (both
ordinary and dotted) run independently from 1 to $N$, the over-lines denote
complex conjugating, $d_b^a$ are the complex numbers satisfying the conditions
$c_a^b d_c^a=\delta^b_c$ ($\delta^b_c$ is the Kronecker symbol), $\det\|c^a_b\|
=\det\|d^a_b\|=1$, and $k$, $l$, $m$, $n$ are nonnegative integers.

Every mapping (7), which possesses the property (8), is called a {\it
Finslerian $N$-spintensor of a valency\/} $\valency{k}{l}{m}{n}$. The addition
and multiplication of such $N$-spintensors are defined in the standard way: if
$\bs{S}$ and $\bs{T}$ have the valency $\valency{k}{l}{m}{n}$ while $\bs{U}$
has the valency $\valency{p}{q}{r}{s}$, then
\begin{align*}
(S+T)_{a_1\!\dots a_m\dot{d}_1\!\dots\dot{d}_n}^{b_1\!\dots b_k\dot{c}_1\!\dots
\dot{c}_l}\!\{\bs{\epsilon}_1,\dots,\bs{\epsilon}_N\}
&=
S_{a_1\!\dots a_m\dot{d}_1\!\dots\dot{d}_n}^{b_1\!\dots b_k\dot{c}_1\!\dots
\dot{c}_l}\!\{\bs{\epsilon}_1,\dots,\bs{\epsilon}_N\}\\
&+
T_{a_1\!\dots a_m\dot{d}_1\!\dots\dot{d}_n}^{b_1\!\dots b_k\dot{c}_1\!\dots
\dot{c}_l}\!\{\bs{\epsilon}_1,\dots,\bs{\epsilon}_N\}
\end{align*}
are the components of the sum $\bs{S}+\bs{T}$ while
\begin{align*}
(S\otimes U)_{a_1\!\dots a_{m+r}\dot{d}_1\!\dots\dot{d}_{n+s}}^{b_1\!\dots
b_{k+p}\dot{c}_1\!\dots\dot{c}_{l+q}}\!\{\bs{\epsilon}_1,\dots,\bs{\epsilon}_N
\}
&=
S_{a_1\!\dots a_m\dot{d}_1\!\dots\dot{d}_n}^{b_1\!\dots b_k\dot{c}_1\!\dots
\dot{c}_l}\!\{\bs{\epsilon}_1,\dots,\bs{\epsilon}_N\}\\
&\times
U_{a_{m+1}\!\dots a_{m+r}\dot{d}_{n+1}\!\dots\dot{d}_{n+s}}^{b_{k+1}\!\dots
b_{k+p}\dot{c}_{l+1}\!\dots\dot{c}_{l+q}}\!\{\bs{\epsilon}_1,\dots,
\bs{\epsilon}_N\}
\end{align*}
are those of the product $\bs{S}\otimes\bs{U}$ with respect to an arbitrary
canonical basis $\{\bs{\epsilon}_1,\dots,\bs{\epsilon}_N\}\in\E(\FS^N)$. Notice
that all Finslerian $N$-spintensors of the valency $\valency{k}{l}{m}{n}$ form
an $N^{k+l+m+n}$-dimensional vector space over $\C\,$. 

Let $\Herm(N)$ be an {\it$N^2$-dimensional vector space over\/}
$\R$ consisting of Finslerian $N$-spintensors $\bs{X}$ of the valency
$\valency{1}{1}{0}{0}$ whose components satisfy the Hermitian symmetry
conditions
$$
X^{b\dot c}\{\bs{\epsilon}_1,\dots,\bs{\epsilon}_N\}=\overline{X^{c\dot b}
\{\bs{\epsilon}_1,\dots,\bs{\epsilon}_N\}}\eqno(9)
$$
for any $\{\bs{\epsilon}_1,\dots,\bs{\epsilon}_N\}\in\E(\FS^N)$. Besides, let
$\{\bs{E}_0,\bs{E}_1,\dots,\bs{E}_{N^2-1}\}$ be a basis in $\Herm(N)$ and $\{
\bs{\epsilon}_1,\bs{\epsilon}_2,\dots,\bs{\epsilon}_N\}$ be a canonical one in
$\FS^N$. With each $\{\bs{\epsilon}_1^\prime,\bs{\epsilon}_2^\prime,\dots,
\bs{\epsilon}_N^\prime\}\in\E(\FS^N)$, we associate a basis $\{\bs{E}_0^\prime,
\bs{E}_1^\prime,\dots,\bs{E}_{N^2-1}^\prime\}$ in $\Herm(N)$ such that
$$
E^{\prime b\dot c}_\alpha\{\bs{\epsilon}_1^\prime,\dots,\bs{\epsilon}_N^\prime
\}=E^{b\dot c}_\alpha,\eqno(10)
$$
where $E^{b\dot c}_\alpha=E^{b\dot c}_\alpha\{\bs{\epsilon}_1,\dots,
\bs{\epsilon}_N\}$ and $\alpha=0,1,\dots,N^2-1$. In other words, (10) defines
the mapping $\{\bs{\epsilon}_1^\prime,\bs{\epsilon}_2^\prime,\dots,
\bs{\epsilon}_N^\prime\}\mapsto\{\bs{E}_0^\prime,\bs{E}_1^\prime,\dots,
\bs{E}_{N^2-1}^\prime\}$ of $\E(\FS^N)$ into the set of all bases in
$\Herm(N)$. However,
$$
E^{\prime b\dot c}_\alpha\{\bs{\epsilon}_1,\dots,\bs{\epsilon}_N\}=
c^b_f\overline{c^{\dot c}_{\dot g}}\,E^{\prime f\dot g}_\alpha
\{\bs{\epsilon}_1^\prime,\dots,\bs{\epsilon}_N^\prime\}\eqno(11)
$$
(compare it with (8)). Due to (10) and (11), we obtain
$$
E^{\prime b\dot c}_\alpha\{\bs{\epsilon}_1,\dots,\bs{\epsilon}_N\}=c^b_f
\overline{c^{\dot c}_{\dot g}}\,E^{f\dot g}_\alpha.\eqno(12)
$$

Let us consider the following expansions
$$
\bs{E}_\alpha^\prime=L(\text{C}_N)_\alpha^\beta\bs{E}_\beta,\eqno(13)
$$
where $L(\text{C}_N)_\alpha^\beta\in\R$ and $\alpha,\beta=0,1,\dots,
N^2-1$. In order to find $L(\text{C}_N)_\alpha^\beta$ as the functions of
$c^a_b$, it is useful to introduce $N^2$ Finslerian $N$-spintensors
$\bs{E}^\alpha$ of the valency $\valency{0}{0}{1}{1}$ such that
$$
\contr(\bs{E}^\alpha\otimes\bs{E}_\beta)=\delta^\alpha_\beta.
\eqno(14)
$$
It is easy to show that $\bs{E}^\alpha$ exist, are unique, and $E^\alpha_{b\dot
c}=\overline{E^\alpha_{c\dot b}}$ with the notation $E^\alpha_{b\dot c}=
E^\alpha_{b\dot c}\{\bs{\epsilon}_1,\dots,\bs{\epsilon}_N\}$. Using (13) and
(14), we can write
$$
L(\text{C}_N)^\alpha_\beta=\contr(\bs{E}^\alpha\otimes\bs{E}_\beta^\prime).
\eqno(15)
$$
On the other hand, (12) implies
$$
\contr(\bs{E}^\alpha\otimes\bs{E}_\beta^\prime)=E^\alpha_{b\dot c}
c^b_f\overline{c^{\dot c}_{\dot g}}E^{f\dot g}_\beta.\eqno(16)
$$
Thus, according to (15) and (16),
$$
L(\text{C}_N)^\alpha_\beta=E^\alpha_{b\dot c}c^b_f\overline{c^{\dot c}_{\dot
g}}E^{f\dot g}_\beta.\eqno(17)
$$

Let $\text{E}^\alpha=\|E^\alpha_{\dot cb}\|$, $\text{E}_\beta=\|E_\beta^{f\dot
g}\|$, and $\text{E}_\beta^\prime=\|E_\beta^{\prime f\dot g}\{\bs{\epsilon}_1,
\dots,\bs{\epsilon}_N\}\|$. Then, it is possible to rewrite (12) and (17) in
the matrix form respectively as
$$
\text{E}_\beta^\prime=\text{C}_N\text{E}_\beta\text{C}_N^+\eqno(18)
$$
and
$$
L(\text{C}_N)^\alpha_\beta=\tr(\text{E}^\alpha\text{C}_N\text{E}_\beta
\text{C}_N^+),\eqno(19)
$$
where the cross denotes Hermitian conjugating. However, it follows from
(13) that $\text{E}_\beta^\prime=L(\text{C}_N)^\gamma_\beta\text{E}_\gamma$.
Therefore,
$$
\text{C}_N\text{E}_\beta\text{C}_N^+=L(\text{C}_N)^\gamma_\beta\text{E}_\gamma.
\eqno(20)
$$ 
Taking into account (19) and (20), we immediately obtain
$$
L(\text{B}_N\text{C}_N)^\alpha_\beta=\tr(\text{E}^\alpha\text{B}_N\text{C}_N
\text{E}_\beta\text{C}_N^+\text{B}_N^+)=L(\text{B}_N)^\alpha_\gamma
L(\text{C}_N)^\gamma_\beta\eqno(21)
$$
for any $\text{B}_N,\text{C}_N\in\SL(N,\C)$.

Let $L(\text{C}_N)=\|L(\text{C}_N)^\alpha_\beta\|$ and $\text{FL}(N^2,\R)=
\{L(\text{C}_N)\mid\text{C}_N\in\SL(N,\C)\}$. In these terms, (21) means that
$\text{FL}(N^2,\R)$ is a group with respect to the matrix multiplication and
the mapping
$$
L\colon\SL(N,\C)\to\text{FL}(N^2,\R),\quad\text{C}_N\mapsto L(\text{C}_N)
\eqno(22)
$$
is a group epimorphism so that, in particular, $L(1_N)=1_{N^2}$ ($1_N$,
$1_{N^2}$ are the identity matrices of the corresponding orders) and
$L(\text{C}_N^{-1})=L(\text{C}_N)^{-1}$. It is easy to prove that the kernel of
the epimorphism (22) has the form
$$
\ker L=\{e^{i\frac{2\pi k}{N}}1_N\mid k=0,1,\dots,N-1\}.\eqno(23)
$$

Let us return to the relations (13). Since both $\{\bs{E}_0,\dots,
\bs{E}_{N^2-1}\}$ and $\{\bs{E}_0^\prime,\dots,\bs{E}_{N^2-1}^\prime\}$ are
bases in $\Herm(N)$, any vector $\bs{X}\in\Herm(N)$ can be expanded in the two
ways
$$
\bs{X}=X^\alpha\bs{E}_\alpha=X^{\prime\beta}\bs{E}^\prime_\beta,\eqno(24)
$$
where $X^\alpha$, $X^{\prime\beta}\in\R$. It is obvious that $X^{\prime\beta}=
L(\text{C}_N^{-1})^\beta_\alpha X^\alpha$. On the other hand, $X^{b\dot c}\{
\bs{\epsilon}_1,\dots,\bs{\epsilon}_N\}=c^b_f\overline{c^{\dot c}_{\dot g}}\,
X^{f\dot g}\{\bs{\epsilon}_1^\prime,\dots,\bs{\epsilon}_N^\prime\}$ or, what is
the same,
$$
\|X^{b\dot c}\{\bs{\epsilon}_1,\dots,\bs{\epsilon}_N\}\|=\text{C}_N\|X^{f\dot
g}\{\bs{\epsilon}_1^\prime,\dots,\bs{\epsilon}_N^\prime\}\|\text{C}_N^+.
\eqno(25)
$$
Remembering that $\det\text{C}_N=1$ and calculating the determinant of (25), we
see that
$$
\det\|X^{b\dot c}\{\bs{\epsilon}_1,\dots,\bs{\epsilon}_N\}\|=\det\|X^{f\dot g}
\{\bs{\epsilon}_1^\prime,\dots,\bs{\epsilon}_N^\prime\}\|\eqno(26)
$$
for {\it any\/} $\{\bs{\epsilon}_1^\prime,\dots,\bs{\epsilon}_N^\prime\}\in
\E(\FS^N)$. Hence, (26) gives an invariant numerical characteristic of the
vector $\bs X$, which is naturally denoted by $\det\bs{X}$. Notice that $\det
\bs{X}=\det\|X^{b\dot c}\{\bs{\epsilon}_1,\dots,\bs{\epsilon}_N\}\|\in\R$ as
it follows from (9).

Thus, without loss of generality, it is possible to calculate $\det\bs{X}$ with
respect to the basis $\{\bs{\epsilon}_1,\dots,\bs{\epsilon}_N\}\in\E(\FS^N)$.
According to (24) and (26), $\det{\bs X}=\det(X^\alpha\text{E}_\alpha)=
\det(X^{\prime\beta}\text{E}^\prime_\beta)$. However, (18) implies
$\det(X^{\prime\beta}\text{E}^\prime_\beta)=\det(\text{C}_N X^{\prime\beta}
\text{E}_\beta\text{C}_N^+)=\det(X^{\prime\beta}\text{E}_\beta)$. Therefore,
$$
\det{\bs X}=\det(X^\alpha\text{E}_\alpha)=\det(X^{\prime\alpha}
\text{E}_\alpha).\eqno(27)
$$
At the same time,
$$
\det(X^\alpha\text{E}_\alpha)=G_{\alpha\beta\dots\gamma}\underbrace{X^\alpha
X^\beta\cdots X^\gamma}_{\text{$N$ multiplicands}},\eqno(28)
$$
where the real coefficients $G_{\alpha\beta\dots\gamma}$ are completely
determined by the choice of the basis $\{\bs{E}_0,\dots,\bs{E}_{N^2-1}\}$ in
$\Herm(N)$. Because of (27) and (28),
$$
\det\bs{X}=G_{\alpha\beta\dots\gamma}X^\alpha X^\beta\cdots X^\gamma=
G_{\alpha\beta\dots\gamma}X^{\prime\alpha}X^{\prime\beta}\cdots
X^{\prime\gamma},\eqno(29)
$$
i.e., $\det\bs{X}$ is {\it forminvariant\/} under transformations of the group
$\text{FL}(N^2,\R)$. Notice that (29) is valid for {\it any\/} basis
$\{\bs{E}_0^\prime,\dots,\bs{E}_{N^2-1}^\prime\}$ whose elements are connected
with those of $\{\bs{E}_0,\dots,\bs{E}_{N^2-1}\}$ by the relations (13).

Denoting $\det\bs{X}$ by $\bs{X}^N$ and using (28), we get (with respect to the
basis $\{\bs{E}_0,\dots,\bs{E}_{N^2-1}\}$)
$$
\bs{X}^N=G_{\alpha\beta\dots\gamma}X^\alpha X^\beta\cdots X^\gamma,\eqno(30)
$$
where $G_{\alpha\beta\dots\gamma}$ are symmetric in all the indices and do not
depend on the choice of any canonical basis in $\FS^N$. Thus, (30) correctly
defines the structure of an {\it $N^2$-dimensional flat Finslerian space\/} on
$\Herm(N)$ so that $\bs{X}^N$ is the $N$-th power of the Finslerian length of
the vector $\bs{X}\in\Herm(N)$ (Finsler, 1918). It should be noted that, in
general, the homogeneous algebraic form (30) is not positive definite.

In the next two sections, we shall illustrate the above formalism by the
simplest examples of Finslerian 2- and 3-spinors.

\section*{3. Finslerian 2-spinors}

Let us consider the case when $N=2$. In this case, the functional (1) is the
usual symplectic scalar multiplication on $\FS^2$. Therefore, $\FS^2$ is
isomorphic to the space $\mathbb{S}^2$ of standard 2-spinors (Penrose and
Rindler, 1986) so that Finslerian 2-spinors are {\it identical\/} to Weyl ones.
Below, we reproduce some essential information on 2-spinors which will be
necessary in the next section of this article.

First of all, for any $\{\bs{\epsilon}_1,\bs{\epsilon}_2\}$,
$\{\bs{\epsilon}^\prime_1,\bs{\epsilon}^\prime_2\}\in\E(\FS^2)$ and $\bs{\xi}=
\xi^a\bs{\epsilon}_a=\xi^{\prime b}\bs{\epsilon}^\prime_b\in\FS^2$, (4) implies
$$
\xi^{\prime a}=d^a_b\xi^b,\eqno(31)
$$
where $\xi^{\prime a}$, $\xi^b\in\C\,$, $c^a_b d^b_c=\delta^a_c$, and $a$, $b$,
$c=1$, $2$. Of course, $\text{C}_2$, $\text{D}_2\in\SL(2,\C)$ and $\text{D}_2=
\text{C}_2^{-1}$ with the notation $\text{C}_2=\|c^a_b\|$, $\text{D}_2=
\|d^a_b\|$. In the same way, (6) gives
$$
[\bs{\xi},\bs{\eta}]=\varepsilon_{ab}\,\xi^a\eta^b=\xi^1\eta^2-\xi^2\eta^1
\eqno(32)
$$
for the scalar product of arbitrary 2-spinors $\bs{\xi}$ and $\bs{\eta}$ with
respect to a basis $\{\bs{\epsilon}_1,\bs{\epsilon}_2\}\in\E(\FS^2)$.

Let us assume
$$
\text{E}^\alpha=\frac{1}{2}\sigma^\alpha,\quad\text{E}_\beta=\sigma_\beta,
\eqno(33)
$$
where $\alpha$, $\beta=0$, 1, 2, 3, $\sigma^\alpha=\sigma_\alpha$, and
$$
\sigma_0=\pmatrix
1&0\\
0&1
\endpmatrix,\ 
\sigma_1=\pmatrix
0&1\\
1&0\endpmatrix,\ 
\sigma_2=\pmatrix
0&-i\\
i&0\endpmatrix,\ 
\sigma_3=\pmatrix
1&0\\
0&-1
\endpmatrix\eqno(34)
$$
are the identity and Pauli matrices. Since $\tr(\sigma^\alpha\sigma_\beta)=2
\delta^\alpha_\beta$, this choice guarantees correctness of (14). It follows
from (13), (21), and (24) that
$$
X^{\prime\alpha}=L(\text{D}_2)^\alpha_\beta X^\beta\eqno(35)
$$
for any 4-vector $\bs{X}\in\Herm(2)$. Using (19), (33), and (34), we obtain
$$
L(\text{D}_2)^\alpha_\beta=\frac{1}{2}\tr(\sigma^\alpha\text{D}_2\sigma_\beta
\text{D}_2^+)\eqno(36)
$$
or, in the explicit form,
\begin{align}
L(\text{D}_2)^0_0&=\frac{1}{2}(
d^1_1\overline{d^{\dot 1}_{\dot 1}}+
d^1_2\overline{d^{\dot 1}_{\dot 2}}+
d^2_1\overline{d^{\dot 2}_{\dot 1}}+
d^2_2\overline{d^{\dot 2}_{\dot 2}}),\notag\\
L(\text{D}_2)^0_1&=\frac{1}{2}(
d^1_1\overline{d^{\dot 1}_{\dot 2}}+
d^2_1\overline{d^{\dot 2}_{\dot 2}}+
d^1_2\overline{d^{\dot 1}_{\dot 1}}+
d^2_2\overline{d^{\dot 2}_{\dot 1}}),\notag\\
L(\text{D}_2)^0_2&=\frac{i}{2}(
d^1_2\overline{d^{\dot 1}_{\dot 1}}+
d^2_2\overline{d^{\dot 2}_{\dot 1}}-
d^1_1\overline{d^{\dot 1}_{\dot 2}}-
d^2_1\overline{d^{\dot 2}_{\dot 2}}),\notag\\
L(\text{D}_2)^0_3&=\frac{1}{2}(
d^1_1\overline{d^{\dot 1}_{\dot 1}}+
d^2_1\overline{d^{\dot 2}_{\dot 1}}-
d^1_2\overline{d^{\dot 1}_{\dot 2}}-
d^2_2\overline{d^{\dot 2}_{\dot 2}}),\notag\\
L(\text{D}_2)^1_0&=\frac{1}{2}(
d^1_1\overline{d^{\dot 2}_{\dot 1}}+
d^2_1\overline{d^{\dot 1}_{\dot 1}}+
d^1_2\overline{d^{\dot 2}_{\dot 2}}+
d^2_2\overline{d^{\dot 1}_{\dot 2}}),\notag\\
L(\text{D}_2)^1_1&=\frac{1}{2}(
d^1_1\overline{d^{\dot 2}_{\dot 2}}+
d^2_1\overline{d^{\dot 1}_{\dot 2}}+
d^1_2\overline{d^{\dot 2}_{\dot 1}}+
d^2_2\overline{d^{\dot 1}_{\dot 1}}),\notag\\
L(\text{D}_2)^1_2&=\frac{i}{2}(
d^1_2\overline{d^{\dot 2}_{\dot 1}}+
d^2_2\overline{d^{\dot 1}_{\dot 1}}-
d^1_1\overline{d^{\dot 2}_{\dot 2}}-
d^2_1\overline{d^{\dot 1}_{\dot 2}}),\notag\\
L(\text{D}_2)^1_3&=\frac{1}{2}(
d^1_1\overline{d^{\dot 2}_{\dot 1}}+
d^2_1\overline{d^{\dot 1}_{\dot 1}}-
d^1_2\overline{d^{\dot 2}_{\dot 2}}-
d^2_2\overline{d^{\dot 1}_{\dot 2}}),\notag\\
L(\text{D}_2)^2_0&=\frac{i}{2}(
d^1_1\overline{d^{\dot 2}_{\dot 1}}-
d^2_1\overline{d^{\dot 1}_{\dot 1}}+
d^1_2\overline{d^{\dot 2}_{\dot 2}}-
d^2_2\overline{d^{\dot 1}_{\dot 2}}),\notag\\
L(\text{D}_2)^2_1&=\frac{i}{2}(
d^1_1\overline{d^{\dot 2}_{\dot 2}}-
d^2_1\overline{d^{\dot 1}_{\dot 2}}+
d^1_2\overline{d^{\dot 2}_{\dot 1}}-
d^2_2\overline{d^{\dot 1}_{\dot 1}}),\notag\\
L(\text{D}_2)^2_2&=\frac{1}{2}(
d^1_1\overline{d^{\dot 2}_{\dot 2}}+
d^2_2\overline{d^{\dot 1}_{\dot 1}}-
d^1_2\overline{d^{\dot 2}_{\dot 1}}-
d^2_1\overline{d^{\dot 1}_{\dot 2}}),\notag\\
L(\text{D}_2)^2_3&=\frac{i}{2}(
d^1_1\overline{d^{\dot 2}_{\dot 1}}-
d^2_1\overline{d^{\dot 1}_{\dot 1}}-
d^1_2\overline{d^{\dot 2}_{\dot 2}}+
d^2_2\overline{d^{\dot 1}_{\dot 2}}),\notag\\
L(\text{D}_2)^3_0&=\frac{1}{2}(
d^1_1\overline{d^{\dot 1}_{\dot 1}}-
d^2_1\overline{d^{\dot 2}_{\dot 1}}+
d^1_2\overline{d^{\dot 1}_{\dot 2}}-
d^2_2\overline{d^{\dot 2}_{\dot 2}}),\notag\\
L(\text{D}_2)^3_1&=\frac{1}{2}(
d^1_1\overline{d^{\dot 1}_{\dot 2}}-
d^2_1\overline{d^{\dot 2}_{\dot 2}}+
d^1_2\overline{d^{\dot 1}_{\dot 1}}-
d^2_2\overline{d^{\dot 2}_{\dot 1}}),\notag\\
L(\text{D}_2)^3_2&=\frac{i}{2}(
d^1_2\overline{d^{\dot 1}_{\dot 1}}-
d^2_2\overline{d^{\dot 2}_{\dot 1}}-
d^1_1\overline{d^{\dot 1}_{\dot 2}}+
d^2_1\overline{d^{\dot 2}_{\dot 2}}),\notag\\
L(\text{D}_2)^3_3&=\frac{1}{2}(
d^1_1\overline{d^{\dot 1}_{\dot 1}}-
d^1_2\overline{d^{\dot 1}_{\dot 2}}-
d^2_1\overline{d^{\dot 2}_{\dot 1}}+
d^2_2\overline{d^{\dot 2}_{\dot 2}}).\tag{36a}
\end{align}
In addition, (28), (30), (33), and (34) imply
$$
\bs{X}^2=G_{\alpha\beta}X^\alpha X^\beta=(X^0)^2-(X^1)^2-(X^2)^2-(X^3)^2.
\eqno(37)
$$
Because of (29) and (37), $\Herm(2)$ is isomorphic to the Minkowski space,
$\text{FL}(4,\R)=\text{O}^\uparrow_+(1,3)$, and (22) coincides with the known
2-to-1 epimorphism $\SL(2,\C)\to\text{O}^\uparrow_+(1,3)$ (Penrose and Rindler,
1986).

Let $\FS^2_\R$ be the realification of $\FS^2$ (see the book (Kostrikin and
Manin, 1989) for the detailed information on the general realification
procedure). Then, $\FS^2_\R$ is a 4-dimensional vector space over $\R$ and its
elements are Majorana 4-spinors. Indeed, setting
$$
\xi^1=\xi_\R^1-i\xi_\R^2,\quad \xi^2=\xi_\R^3-i\xi_\R^4,\quad
\xi^{\prime 1}=\xi_\R^{\prime 1}-i\xi_\R^{\prime 2},\quad
\xi^{\prime 2}=\xi_\R^{\prime 3}-i\xi_\R^{\prime 4},\eqno(38)
$$
we obtain
\begin{align}
\bs{\xi}&=
\xi^a\bs{\epsilon}_a=
\xi_\R^1 \bs{\epsilon}_1-
\xi_\R^2 i\bs{\epsilon}_1+
\xi_\R^3 \bs{\epsilon}_2-
\xi_\R^4 i\bs{\epsilon}_2,\notag\\
\bs{\xi}&=
\xi^{\prime b}\bs{\epsilon}^\prime_b=
\xi_\R^{\prime 1} \bs{\epsilon}^\prime_1-
\xi_\R^{\prime 2}i\bs{\epsilon}^\prime_1+
\xi_\R^{\prime 3} \bs{\epsilon}^\prime_2-
\xi_\R^{\prime 4}i\bs{\epsilon}^\prime_2\tag{39}
\end{align}
for any $\{\bs{\epsilon}_1,\bs{\epsilon}_2\}$, $\{\bs{\epsilon}^\prime_1,
\bs{\epsilon}^\prime_2\}\in\E(\FS^2)$ and $\bs{\xi}\in\FS^2$; here $\xi_\R^i$,
$\xi_\R^{\prime j}\in\R$ ($i$, $j=1$, 2, 3, 4). It follows from (39) that
$\{\bs{\epsilon}_1,-i\bs{\epsilon}_1,\bs{\epsilon}_2,-i\bs{\epsilon}_2\}$ and
$\{\bs{\epsilon}^\prime_1,-i\bs{\epsilon}^\prime_1,\bs{\epsilon}^\prime_2,-i
\bs{\epsilon}^\prime_2\}$ are bases in $\FS^2_\R$. Moreover, the substitution
of (38) into (31) provides
$$
\xi_\R^{\prime i}=M(\text{D}_2)^i_j\xi_\R^j,\eqno(40)
$$
where $M(\text{D}_2)^i_j\in\R$ and have the form
\begin{align}
M(\text{D}_2)^1_1&=\frac{1}{2}(\overline{d^1_1}+d^1_1),\quad
M(\text{D}_2)^3_1=\frac{1}{2}(\overline{d^2_1}+d^2_1),\notag\\
M(\text{D}_2)^1_2&=\frac{i}{2}(\overline{d^1_1}-d^1_1),\quad
M(\text{D}_2)^3_2=\frac{i}{2}(\overline{d^2_1}-d^2_1),\notag\\
M(\text{D}_2)^1_3&=\frac{1}{2}(\overline{d^1_2}+d^1_2),\quad
M(\text{D}_2)^3_3=\frac{1}{2}(\overline{d^2_2}+d^2_2),\notag\\
M(\text{D}_2)^1_4&=\frac{i}{2}(\overline{d^1_2}-d^1_2),\quad
M(\text{D}_2)^3_4=\frac{i}{2}(\overline{d^2_2}-d^2_2),\notag\\
M(\text{D}_2)^2_1&=\frac{i}{2}(d^1_1-\overline{d^1_1}),\quad
M(\text{D}_2)^4_1=\frac{i}{2}(d^2_1-\overline{d^2_1}),\notag\\
M(\text{D}_2)^2_2&=\frac{1}{2}(d^1_1+\overline{d^1_1}),\quad
M(\text{D}_2)^4_2=\frac{1}{2}(d^2_1+\overline{d^2_1}),\notag\\
M(\text{D}_2)^2_3&=\frac{i}{2}(d^1_2-\overline{d^1_2}),\quad
M(\text{D}_2)^4_3=\frac{i}{2}(d^2_2-\overline{d^2_2}),\notag\\
M(\text{D}_2)^2_4&=\frac{1}{2}(d^1_2+\overline{d^1_2}),\quad
M(\text{D}_2)^4_4=\frac{1}{2}(d^2_2+\overline{d^2_2}).\tag{41}
\end{align}
It is evident that the matrix group $\text{Maj}(4)=\{\|M(\text{D}_2)^i_j\|\mid
\text{D}_2\in\SL(2,\C)\}$ is isomorphic to $\SL(2,\C)$. Finally, using $\eta^1=
\eta_\R^1-i\eta_\R^2$, $\eta^2=\eta_\R^3-i\eta_\R^4$, and (38), we can rewrite
(32) as $[\bs{\xi},\bs{\eta}]=\overline{\xi}\gamma^5\eta-i\overline{\xi}\eta$,
where $\xi=(\xi^1_\R,\xi^2_\R,\xi^3_\R,\xi^4_\R)^\top$ and $\eta=(\eta^1_\R,
\eta^2_\R,\eta^3_\R,\eta^4_\R)^\top$ are column matrices, the ``$\scriptstyle
\top$'' mark denotes the matrix transposition, $\overline{\xi}=\xi^\top
\gamma^0$ is a row matrix, and
\begin{align}
&\gamma^0=
\pmatrix
0&0&i&0\\
0&0&0&-i\\
-i&0&0&0\\
0&i&0&0
\endpmatrix,\
\gamma^1=
\pmatrix
i&0&0&0\\
0&-i&0&0\\
0&0&-i&0\\
0&0&0&i
\endpmatrix,\
\gamma^2=
\pmatrix
0&i&0&0\\
i&0&0&0\\
0&0&0&i\\
0&0&i&0
\endpmatrix,\notag\\
&\gamma^3=
\pmatrix
0&0&-i&0\\
0&0&0&i\\
-i&0&0&0\\
0&i&0&0
\endpmatrix,\
\gamma^5=\gamma^0\gamma^1\gamma^2\gamma^3=
\pmatrix
0&-1&0&0\\
1&0&0&0\\
0&0&0&-1\\
0&0&1&0
\endpmatrix\tag{42}
\end{align}
are Dirac matrices in a Majorana representation (Majorana, 1937) which satisfy
the standard conditions $\gamma^\alpha\gamma^\beta+\gamma^\beta\gamma^\alpha=2
g^{\alpha\beta}$ with $(g^{\alpha\beta})=\text{diag}\,(1, -1, -1,$ $-1)$.

\section*{4. Finslerian 3-spinors}

In this section, we consider the nontrivial case of Finslerian $N$-spinors when
$N=3$. Besides, the algebraic structure of the group $\text{FL}(9,\R)$ is also
described here.

Let us begin with the following remark. For any $\{\bs{\epsilon}_1,
\bs{\epsilon}_2,\bs{\epsilon}_3\}$, $\{\bs{\epsilon}^\prime_1,
\bs{\epsilon}^\prime_2,\bs{\epsilon}^\prime_3\}\in\E(\FS^3)$ and $\bs{\xi}=
\xi^a\bs{\epsilon}_a=\xi^{\prime b}\bs{\epsilon}^\prime_b\in\FS^3$, (4) implies
$\xi^{\prime a}=d^a_b\xi^b$, where $\xi^{\prime a}$, $\xi^b\in\C\,$, $c^a_b
d^b_c=\delta^a_c$, and $a$, $b$, $c=1$, $2$, $3$. It is clear that $\text{C}_3$,
$\text{D}_3\in\SL(3,\C)$ and $\text{D}_3=\text{C}_3^{-1}$ with the notation
$\text{C}_3=\|c^a_b\|$, $\text{D}_3=\|d^a_b\|$. In the same way, (6) gives
$[\bs{\xi},\bs{\eta},\bs{\zeta}]=\varepsilon_{abc}\,\xi^a\eta^b\zeta^c$ for the
scalar 3-product of arbitrary Finslerian 3-spinors $\bs{\xi}$, $\bs{\eta}$, and
$\bs{\zeta}$ with respect to a basis $\{\bs{\epsilon}_1,\bs{\epsilon}_2,
\bs{\epsilon}_3\}\in\E(\FS^3)$.

By analogy with the previous section, we set
$$
\text{E}^A=\frac{1}{2}\lambda^A,\quad\text{E}_B=\lambda_B,\eqno(43)
$$
where $A,B=0,1,\dots,8$, $\lambda^A=\lambda_A$ ($A\ne8$), $\lambda^8=2
\lambda_8$, and
\begin{align}
&\lambda_0=
\pmatrix
1&0&0\\
0&1&0\\
0&0&0
\endpmatrix,\quad\phantom{-}
\lambda_1=
\pmatrix
0&1&0\\
1&0&0\\
0&0&0
\endpmatrix,\quad\phantom{-}
\lambda_2=
\pmatrix
0&-i&0\\
i&0&0\\
0&0&0
\endpmatrix,\notag\\
&\lambda_3=
\pmatrix
1&0&0\\
0&-1&0\\
0&0&0
\endpmatrix,\quad
\lambda_4=
\pmatrix
0&0&1\\
0&0&0\\
1&0&0
\endpmatrix,\quad\phantom{-}
\lambda_5=
\pmatrix
0&0&-i\\
0&0&0\\
i&0&0
\endpmatrix,\notag\\
&\lambda_6=
\pmatrix
0&0&0\\
0&0&1\\
0&1&0
\endpmatrix,\quad\phantom{-} 
\lambda_7=
\pmatrix
0&0&0\\
0&0&-i\\
0&i&0
\endpmatrix,\quad
\lambda_8=
\pmatrix
0&0&0\\
0&0&0\\
0&0&1
\endpmatrix\tag{44}
\end{align}
($\lambda_1,\lambda_2,\dots,\lambda_7$ coincide with the corresponding
Gell-Mann matrices). Since $\tr(\lambda^A\lambda_B)=2\delta^A_B$, the choice
(43) guarantees correctness of (14). It follows from (13), (21), and (24) that
$$
X^{\prime A}=L(\text{D}_3)^A_B X^B\eqno(45)
$$
for any 9-vector $\bs{X}\in\Herm(3)$. Using (19), (43), and (44), we obtain
$$
L(\text{D}_3)^A_B=\frac{1}{2}\tr(\lambda^A\text{D}_3\lambda_B\text{D}_3^+).
\eqno(46)
$$
In addition, (28), (30), (43), and (44) imply
\begin{align}
\bs{X}^3&=G_{AB\varGamma}X^A X^B X^\varGamma=
[(X^0)^2-(X^1)^2-(X^2)^2-(X^3)^2]X^8\notag\\
&-X^0[(X^4)^2+(X^5)^2+(X^6)^2+(X^7)^2]\notag\\
&+2X^1[X^4X^6+X^5X^7]+2X^2[X^5X^6-X^4X^7]\notag\\
&+X^3[(X^4)^2+(X^5)^2-(X^6)^2-(X^7)^2].\tag{47}
\end{align}
Because of (29), the Finslerian ``scalar cube'' (47) is forminvariant under the
transformations (45), (46) of the group $\text{FL}(9,\R)$.

It is more or less clear that any matrix $\widehat{\text{D}}_3\in\SL(3,\C)$
with ${\hat d}^3_3\ne 0$ can be represented in a form of the product
$$
\widehat{\text{D}}_3=\text{D}_3^{(1)}\text{D}_3^{(2)}\text{D}_3^{(3)}
\text{D}_3^{(4)},\eqno(48)
$$
where
\begin{align}
&\text{D}^{(1)}_3=
\pmatrix
d^1_1&d^1_2&0\\
d^2_1&d^2_2&0\\
0&0&1
\endpmatrix,\quad
\text{D}_3^{(2)}=
\pmatrix 
1&0&d^1_3\\
0&1&d^2_3\\
0&0&1
\endpmatrix,\notag\\
&\text{D}_3^{(3)}=
\pmatrix
1&0&0\\
0&1&0\\
d^3_1&d^3_2&1
\endpmatrix,\quad
\text{D}_3^{(4)}=
\pmatrix
d&0&0\\
0&d&0\\
0&0&d^{-2}
\endpmatrix\tag{49}
\end{align}
are $\SL(3,\C)$ matrices too. Due to (21), (48), and (49), we obtain the
decomposition
$$
L(\widehat{\text{D}}_3)=L(\text{D}_3^{(1)})L(\text{D}_3^{(2)})
L(\text{D}_3^{(3)})L(\text{D}_3^{(4)})\eqno(50)
$$
of the corresponding $\text{FL}(9,\R)$ matrix $L(\widehat{\text{D}}_3)$. Thus,
(50) reduces a general $\text{FL}(9,\R)$ transformation $X^{\prime A}=
L(\widehat{\text{D}}_3)^A_B X^B$ to a composition of four simpler ones induced
by the matrices (49). These $\text{FL}(9,\R)$ transformations will be explicitly
described below.

Let $\xi^i_\R=X^{3+i}$ ($i=1,2,3,4$). Then, with the help of (45) and (46), the
$\text{FL}(9,\R)$ transformation $X^{\prime A}=L(\text{D}_3^{(1)})^A_B X^B$ is
written in the following form
\begin{align}
X^{\prime\alpha}&=L(\text{D}_2)^\alpha_\beta X^\beta,\notag\\
\xi^{\prime i}_\R&=M(\text{D}_2)^i_j\xi^j_\R,\notag\\
X^{\prime 8}&=X^8,\tag{51}
\end{align}
where $\alpha,\beta=0,1,2,3$ and $i,j=1,2,3,4$. It is easy to see that the first
line of (51) coincides with the Lorentz transformation (35), (36), (36a) of a
4-vector $X^\alpha$, while the second line is the transformation (40), (41) of
a Majorana 4-spinor $\xi^i_\R$. Therefore, the transformations (51) form a
6-parametric non-Abelian subgroup of $\text{FL}(9,\R)$.

Let $d^1_3=\varepsilon^1-i\varepsilon^2$, $d^2_3=\varepsilon^3-i\varepsilon^4$
be a parametrization of the complex matrix $\text{D}_3^{(2)}$ from (49).
Introducing the real column matrices $\varepsilon=(\varepsilon^1,\varepsilon^2,
\varepsilon^3,\varepsilon^4)^\top$, $\xi=(X^4,X^5,X^6,X^7)^\top$ and using (42),
(45), (46), we can write the $\text{FL}(9,\R)$ transformation $X^{\prime A}=
L(\text{D}_3^{(2)})^A_B X^B$ as
\begin{align}
X^{\prime\alpha}&=X^\alpha+\overline{\varepsilon}\gamma^\alpha\xi+\tfrac{1}{2}
\overline{\varepsilon}\gamma^\alpha\varepsilon X^8,\notag\\
\xi^\prime&=\xi+\varepsilon X^8,\notag\\
X^{\prime 8}&=X^8,\tag{52}
\end{align}
where $\alpha=0,1,2,3$ and $\overline{\varepsilon}=\varepsilon^\top\gamma^0$.
Since $\varepsilon^1$, $\varepsilon^2$, $\varepsilon^3$, $\varepsilon^4\in\R$,
the transformations (52) form a 4-parametric Abelian subgroup of $\text{FL}(9,
\R)$.

Let $d_1^3=\varkappa^3-i\varkappa^4$, $d_2^3=-\varkappa^1+i\varkappa^2$ be a
parametrization of the complex matrix $\text{D}_3^{(3)}$ from (49). Introducing
the real column matrices $\varkappa=(\varkappa^1$, $\varkappa^2$, $\varkappa^3$,
$\varkappa^4)^\top$, $\xi=(X^4,X^5,X^6,X^7)^\top$ and using (42), (45), (46), we
write the $\text{FL}(9,\R)$ transformation $X^{\prime A}=L(\text{D}_3^{(3)})^A_B
X^B$ as
\begin{align}
X^{\prime\alpha}&=X^\alpha,\notag\\
\xi^\prime&=-ig_{\alpha\beta}\gamma^\alpha\varkappa X^\beta+\xi,\notag\\
X^{\prime 8}&=g_{\alpha\beta}\overline{\varkappa}\gamma^\alpha\varkappa X^\beta
+2i\overline{\varkappa}\xi+X^8,\tag{53}
\end{align}
where $\alpha,\beta=0,1,2,3$, $\overline{\varkappa}=\varkappa^\top\gamma^0$, and
$(g_{\alpha\beta})=\text{diag}(1,-1,-1,-1)$. Since $\varkappa^1$, $\varkappa^2$,
$\varkappa^3$, $\varkappa^4\in\R$, the transformations (53) form a 4-parametric
Abelian subgroup of $\text{FL}(9,\R)$.

Let $d=|d|e^{i\varphi}\ne 0$ be a parametrization of the complex matrix
$\text{D}_3^{(4)}$ from (49). Using (45) and (46), we represent the $\text{FL}
(9,\R)$ transformation $X^{\prime A}=L(\text{D}_3^{(4)})^A_B X^B$ in the
following form
\begin{align}
X^{\prime\alpha}&=|d|^2 X^\alpha,\notag\\
\pmatrix
X^{\prime 4}\\
X^{\prime 5}
\endpmatrix
&=|d|^{-1}
\pmatrix
\cos 3\varphi&\sin 3\varphi\\
-\sin 3\varphi&\cos 3\varphi\\
\endpmatrix
\pmatrix
X^4\\
X^5
\endpmatrix,\notag\\
\pmatrix
X^{\prime 6}\\
X^{\prime 7}
\endpmatrix
&=|d|^{-1}
\pmatrix
\cos 3\varphi&\sin 3\varphi\\
-\sin 3\varphi&\cos 3\varphi
\endpmatrix
\pmatrix
X^6\\
X^7
\endpmatrix,\notag\\
X^{\prime 8}&=|d|^{-4}X^8,\tag{54}
\end{align}
where $\alpha=0,1,2,3$. Since $|d|>0$ and $\varphi\in\R$, the transformations
(54) form a 2-parametric Abelian subgroup of $\text{FL}(9,\R)$.

Thus, all of the four $\text{FL}(9,\R)$ transformations corresponding to the
matrices of the decomposition (50) have been explicitly described in (51), (52),
(53), and (54). Finally, with the above notation, it is possible to rewrite (47)
as $\bs{X}^3=g_{\alpha\beta}X^\alpha X^\beta X^8-g_{\alpha\beta}X^\alpha
\overline{\xi}\gamma^\beta\xi$.

\section*{5. Conclusion}

In the present article, we have considered algebraic aspects of the Finslerian
$N$-spinor theory. We formulated the general definitions of a Finslerian
$N$-spinor and Finslerian $N$-spintensor of an arbitrary valency. It was shown
that Finslerian $N$-spintensors of the valency $\valency{1}{1}{0}{0}$ were
closely associated with the $N^2$-dimensional flat Finslerian space $\Herm(N)$.
The metric on $\Herm(N)$ was characterized by the homogeneous algebraic form
(30) of the $N$-th power. We also constructed the generalization (22) of the
well known epimorphism $\SL(2,\C)\to\text{O}_+^\uparrow(1,3)$ and found that
its kernel consisted of the $N$ scalar matrices (23). In particular, it turned
out that Finslerian 2-spinors coincided with standard Weyl spinors. In this
connection, we recalled some essential information on Majorana 4-spinors as
well. Finally, we considered properties of Finslerian 3-spinors and described
the algebraic structure of the group $\text{FL}(9,\R)$.

When this article had already been written, we knew about the remarkable works
(Finkelstein, 1986; Finkelstein et al., 1986) in which {\it hyperspinors\/} and
some of their properties were considered. David Finkelstein's hyperspinors
actually coincide with Finslerian $N$-spinors for which we have developed the
detailed algebraic theory above. We are grateful to Andrei Galiautdinov for
attracting our attention to the works on hyperspinors.

\section*{References}

\begin{description}
\item[]Brauer, R. and Weyl, H. (1935). {\it American Journal of Mathematics},
{\bf 57}, 425--449.
\item[]Cartan, \'E. (1913). {\it Bulletin de la Soci\'et\'e Math\'ematique de
France}, {\bf 41}, 53--96.
\item[]Cartan, \'E. (1938). {\it Le\c{c}ons sur la th\'eorie des spineurs}.
Actualit\'es scientifiques et industrielles, Paris. 
\item[]Dirac, P. A. M. (1928). {\it Proceedings of the Royal Society (London)},
{\bf A117}, 610--624.
\item[]Finkelstein, D. (1986). {\it Physical Review Letters}, {\bf 56},
1532--1533.
\item[]Finkelstein, D., Finkelstein, S. R., and Holm, C. (1986). {\it
International Journal of Theoretical Physics}, {\bf 25}, 441--463.
\item[]Finsler, P. (1918). {\it \"Uber Kurven und Fl\"achen in allgemeinen
R\"aumen}. Dissertation, G\"ottingen.
\item[]Kostrikin, A. I. and Manin, Yu. I. (1989). {\it Linear algebra and
geometry}. Gordon and Breach, New York. 
\item[]Majorana, E. (1937). {\it Nuovo Cimento}, {\bf 14}, 171--184.
\item[]Pauli, W. (1927). {\it Zeitschrift f\"{u}r Physik}, {\bf 43}, 601--623.
\item[]Penrose, R. and Rindler, W. (1986). {\it Spinors and space-time}.
Cambridge University Press, Cambridge.
\item[]Solov'yov, A. V. (1996). {\it $N$-spinor calculus in the relational
theory of space-time}. Doctor of Philosophy thesis, Moscow (in Russian).
\item[]Vladimirov, Yu. S. (1996). {\it The relational theory of space-time and
interactions}. Moscow University Press, Moscow (in Russian).
\end{description}

\end{document}